\documentclass[]{spie}  

\usepackage{amsmath,amsfonts,amssymb}
\usepackage{graphicx}
\usepackage{subcaption}
\usepackage{float}
\usepackage[colorlinks=true, allcolors=blue]{hyperref}
\usepackage[toc,page]{appendix}
\usepackage{authblk}

\title{Automatic Cerebral Vessel Extraction in TOF-MRA Using Deep Learning}

\begin{document} 
	
\author[1]{V. de Vos}
\author[2]{K.M. Timmins}
\author[2]{I.C. van der Schaaf}
\author[2]{Y. Ruigrok}
\author[2]{B.K. Velthuis}
\author[2]{H.J. Kuijf}
\affil[1]{Eindhoven University of Technology, The Netherlands}
\affil[2]{University Medical Center Utrecht, The Netherlands}

\pagestyle{empty} 
\setcounter{page}{301} 
 
\maketitle

\begin{abstract}
Deep learning approaches may help radiologists in the early diagnosis and timely treatment of cerebrovascular diseases. Accurate cerebral vessel segmentation of Time-of-Flight Magnetic Resonance Angiographs (TOF-MRAs) is an essential step in this process. This study investigates deep learning approaches for automatic, fast and accurate cerebrovascular segmentation for TOF-MRAs.

The performance of several data augmentation and selection methods for training a 2D and 3D U-Net for vessel segmentation was investigated in five experiments: a) without augmentation, b) Gaussian blur, c) rotation and flipping, d) Gaussian blur, rotation and flipping and e) different input patch sizes. All experiments were performed by patch-training both a 2D and 3D U-Net and predicted on a test set of MRAs. Ground truth was manually defined using an interactive threshold and region growing method. The performance was evaluated using the Dice Similarity Coefficient (DSC), Modified Hausdorff Distance and Volumetric Similarity, between the predicted images and the interactively defined ground truth.

The segmentation performance of all trained networks on the test set was found to be good, with DSC scores ranging from 0.72 to 0.83. Both the 2D and 3D U-Net had the best segmentation performance with Gaussian blur, rotation and flipping compared to other experiments without augmentation or only one of those augmentation techniques. Additionally, training on larger patches or slices gave optimal segmentation results.

In conclusion, vessel segmentation can be optimally performed on TOF-MRAs using a trained 3D U-Net on larger patches, where data augmentation including Gaussian blur, rotation and flipping was performed on the training data.
 
\end{abstract}


\section{Introduction} \label{Introduction}
Stroke, including ischemic and hemorrhagic stroke and aneurysmal subarachnoid hemorrhage, is a major cause of death and disability worldwide with more than six million deaths in 2015 \cite{Phellan2007}. In some cases it can be caused by abnormalities of the intracranial arteries including stenosis, intracranial aneurysms and other vascular malformations. The incidence is even increasing because of the increasing population ages \cite{Katan2018,Phellan2007}. 

For an early diagnosis and timely treatment of various cerebrovascular diseases, detailed information about the vasculature might aid a radiologist in decision making. This information could be obtained from cerebrovascular segmentations, where the blood vessels are extracted from the images. This will allow for quantitative analysis of the vasculature, as well as better (3D) visualization \cite{Phellan2007, Frangi1998, Gan2005}. Currently, use of such segmentations is not common practice, because this often requires manual segmentation; a difficult and time-consuming procedure, which is prone to inter- and intra-rater variability \cite{Phellan2007, Livne2019}. Automatic vessel extraction methods could overcome this issue, including methods as Markov random fields \cite{Fang2011}, multi scale filtering \cite{Frangi1998}, deformable models \cite{McInerney1997}, hybrid methods \cite{Chen2003} and deep learning \cite{Phellan2007, Livne2019}. Such methods create a 3D vascular model for every patient, which can be useful to find vessel abnormalities \cite{Gan2005}. In a study of Gan et al. (2005), an automatic vessel segmentation method based on maximum intensity projections (MIP) was presented. This method compiled the vessel segmentation iteratively by using the segmentation of the MIP images along a fixed direction. The MIP images were segmented with a finite mixture model (FMM) and expectation maximization (EM) algorithm. Once the images were segmented along the individual axes, the results were combined \cite{Gan2005}. In addition, a study of Phellan et al. (2017) proposed a deep Convolutional Neural Network (CNN) to automatically segment the vessels in TOF-MRA images of healthy subjects. Experiments were performed with a varying number of images for training the CNN and cross validation was used to test the generalization of the model. The ground truth was obtained by manual annotated image patches extracted in the axial, coronal and sagittal directions \cite{Phellan2007}. 

This study provides an automatic vessel segmentation method by training and evaluating a CNN with U-Net architecture \cite{Ronneberger2015}, which is one of the most promising deep learning networks for segmentation tasks. To evaluate the performance of this network, different experiments were performed to compare a 2D and 3D U-Net architecture with several training data augmentation and selection methods.

\section{Materials and Methods} \label{Materials and Methods}

\subsection{Dataset} \label{Dataset}
\begin{figure}[H]
	\centering
	\includegraphics[width=0.4\textwidth]{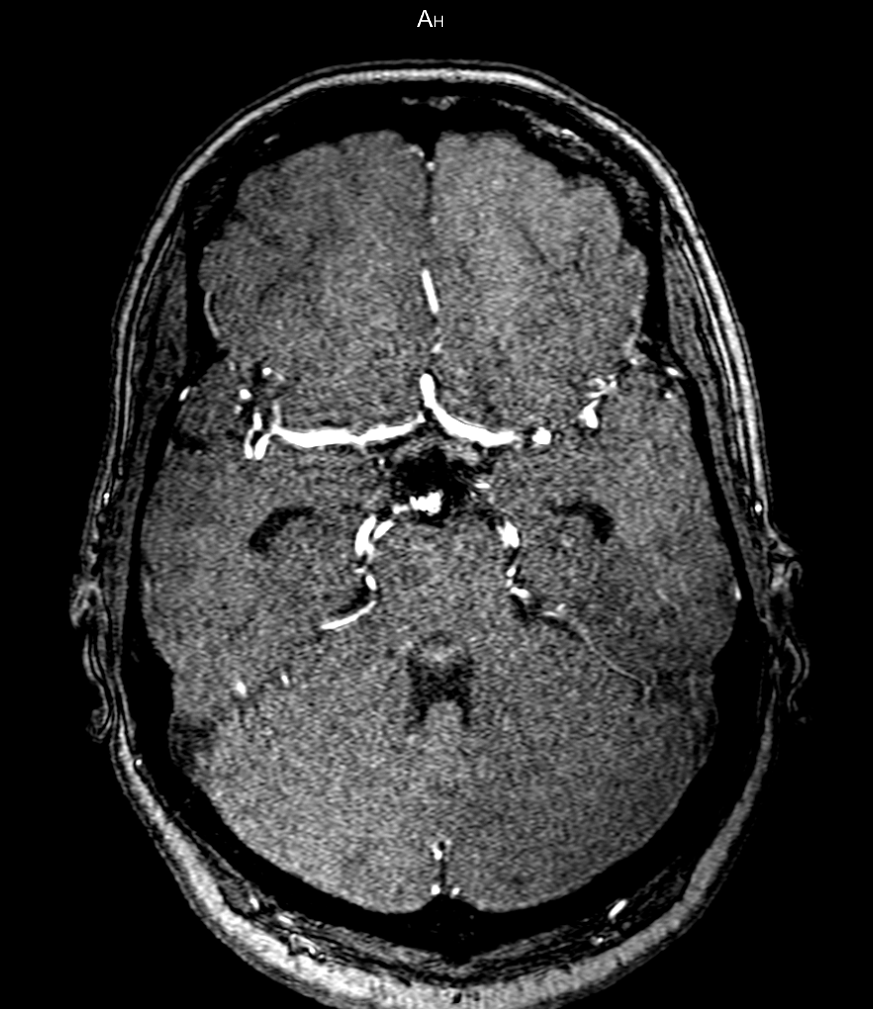}
	\vspace*{2mm}
	\caption{Example slice in the transverse plane of a TOF-MRA.}
	\label{fig:scan}
\end{figure}

The data used in this study included 69 patients with unruptured aneurysms scanned in the University Medical Center Utrecht, the Netherlands. All patients underwent a 3D TOF-MRA scan in the period between 2004 and 2012  and were scanned twice, a baseline scan and a follow up scan. An example of one slice of a TOF-MRA is shown in Figure \ref{fig:scan}. Overall, the slice thickness ranged from 0.4 to 0.7 mm and the in-plane voxel size ranged from 0.195x0.195 to 0.586x0.586 mm. 

\subsection{Pre-processing} \label{Pre-processing}
Before segmenting the vascular structure, the images in the dataset, as described in section \ref{Dataset}, were preprocessed by using N4 bias field inhomogeneity correction \cite{Tustison2010, ANTs} and Z-score normalization \cite{Ellingson2012}. 

The dataset did not contain delineations of the brain vasculature. To acquire the labelled ground truth data for vessel segmentation, interactive vessel segmentation was performed. First, the image was interactively thresholded by using histogram-based thresholding in which the user can choose the image specific intensity percentage at which the threshold was determined. The threshold of all images was chosen between 95\% and 99\% of the maximum image intensity. The resulting thresholded image was used to define seed points for region growing. The resulting labels were manually checked for accuracy and corrected as required. The interactive vessel segmentation was performed in MevisLab (version 3.2) \cite{Ritter2011}.

\subsection{Network} \label{Network}
Both a 2D and 3D fully convolutional neural network with U-Net architecture \cite{Ronneberger2015} were trained on randomly selected and augmented patches from TOF-MRA images. For the 2D network, the input patches had a size of 64x64 voxels and for the 3D network a size of 16x16x16 voxels in order to train on the same number of voxels per patch in 2D and 3D. The same patches were used for all the experiments. 

A balanced number of patches from vessel (80\%) and non-vessel (20\%) regions were used for training. The selection of patches was based on the center voxel of each patch. When this voxel was labelled as vessel in the ground truth image, the patch was categorized as a patch containing vessels and otherwise it was categorized as non-vessel patch.

Finally, both the 2D and 3D network were optimized using a dice loss function, Adam optimizer and a learning rate of $1\times10^{-4}$.

\subsection{Experiments} \label{Experiments}
For both the 2D and 3D architectures (190.396 trainable parameters), five experiments were compared. In all experiments, the same MRAs were used for training (n = 84, 64\%), validation (n = 21, 16\%) and test (n = 26, 20\%). The first experiment, (a), was performed without applying any augmentation technique to the training data. Next, three experiments were performed by training the networks with the patches with different augmentation techniques: b) Gaussian blurring, c) rotation and flipping and d) both Gaussian blurring and rotation and flipping. The fifth experiment, (e), was performed by training the networks with full slices instead of patches for 2D and training the 3D network with larger patches (64x64x64 voxels) with all augmentation techniques mentioned before.

The resulting trained networks were used to segment the blood vessels in the pre-processed test set of MRAs. Voxels with a probability larger than 0.7 were assumed to be inside a vessel. 

Post-processing was performed using connected component analysis in which regions with less than 200 voxels were eliminated from the segmentation. 

\subsection{Evaluation metrics} \label{Evaluation metrics}
To evaluate and compare the performances of the different experiments, the Dice Similarity Coefficient (DSC) \cite{Toennies-dsc, Taha2015}, Modified Hausdorff Distance (MHD) \cite{Toennies-dsc, Taha2015} and Volumetric Similarity (VS) \cite{Taha2015} between the predicted segmentation and the generated ground truth segmentation for each MRA were determined. 

The DSC was used to evaluate the overlap between the ground truth and predicted segmentation. However, the DSC is limited for the evaluation of the vessel segmentations as vessels are narrow and elongated. For this reason, segmentation errors can quickly lead to a loss of overlap. Therefore, a distance metric was also used for evaluation \cite{Livne2019}. A commonly used distance metric is the Hausdorff Distance (HD). However, this measure is very sensitive to outliers, which are common in medical segmentations. For this reason, the Modified Hausdorff Distance (MHD) was used, which is not based on the maximum distance between points but on a defined percentile (95\%) of the distance between boundary points \cite{Toennies-dsc, Taha2015}. Finally, the VS was used to compare the segmented volumes without taking into account the location or overlap of the segmentations.

A Wilcoxon signed-rank test was performed to compare the results achieved by the different experiments. This test was performed with the goal of determining whether there is a difference between the evaluation metrics of the experiments \cite{Whitley2002}. Python version 3.7.6 with the SciPy library was used to perform this test.

\section{Results} \label{Results}
Tables \ref{tab:2DUnet} and \ref{tab:3DUnet} show the average resulting numerical results expressing the performance of the experiments in both 2D and 3D, respectively. It can be observed that the segmentation performance of all trained networks in both 2D and 3D was good with all mean DSC scores larger than 0.70.

\begin{table}[H]
\caption{Segmentation metrics for the test set for the proposed augmentation techniques and the use of patches or slices for the training of the U-Net. Values are provided as the mean $\pm$ the standard deviation. The size in voxels of the patches used for the different experiments are indicated between the brackets. (a) 2D U-Net, (b) 3D U-Net.}
\begin{subtable}{\textwidth}
	\caption{2D U-Net}
	\begin{tabular}{|l|l|l|l|l|l|}
		\hline
		& \emph{\textbf{2D U-Net}} & \textbf{Augmentation} & \textbf{DSC} & \textbf{MHD [mm]} & \textbf{VS} \\ \hline
		a & Patches & None & 0.74 $\pm$ 0.17 & 47.6 $\pm$ 40.4 & 0.74 $\pm$ 0.18 \\ 
		& (64x64) & & & & \\ \hline
		b & Patches & Gaussian blur & 0.81 $\pm$ 0.12 & 41.6 $\pm$ 42.5 & 0.83 $\pm$ 0.13 \\ 
		& (64x64) & & & & \\ \hline
		c & Patches & Rotation and flipping & 0.80 $\pm$ 0.14 & 35.8 $\pm$ 39.0 & 0.84 $\pm$ 0.16 \\ 
		& (64x64) & & & & \\ \hline
		d & Patches & Gaussian blur, rotation and flipping & 0.82 $\pm$ 0.15 & 34.1 $\pm$ 42.5 & 0.85 $\pm$ 0.17 \\
		& (64x64) & & & & \\ \hline
		e & Slices & Gaussian blur, rotation and flipping & 0.83 $\pm$ 0.14 & 28.0 $\pm$ 37.0 & 0.85 $\pm$ 0.16 \\ 
		& & & & & \\ \hline
	\end{tabular}
	\label{tab:2DUnet}
\end{subtable}

\vspace*{0.2 cm}

\begin{subtable}{\textwidth}
	\caption{3D U-Net}
	\begin{tabular}{|l|l|l|l|l|l|}
		\hline
		& \emph{\textbf{3D U-Net}} & \textbf{Augmentation} & \textbf{DSC} & \textbf{MHD [mm]} & \textbf{VS} \\ \hline
		a & Patches & None & 0.72 $\pm$ 0.15 & 81.3 $\pm$ 57.0 & 0.78 $\pm$ 0.17 \\ 
		& (16x16x16) & & & & \\ \hline
		b & Patches & Gaussian blur & 0.76 $\pm$ 0.15 & 27.5 $\pm$ 30.3 & 0.77 $\pm$ 0.16 \\ 
		& (16x16x16) & & & & \\ \hline
		c & Patches & Rotation and flipping & 0.79 $\pm$ 0.12 & 33.7 $\pm$ 33.1 & 0.85 $\pm$ 0.15 \\
		& (16x16x16) & & & & \\ \hline
		d & Patches & Gaussian blur, rotation and flipping & 0.81 $\pm$ 0.12 & 36.6 $\pm$ 37.0 & 0.85 $\pm$ 0.15 \\
		& (16x16x16) & & & & \\ \hline
		e & Patches & Gaussian blur, rotation and flipping & 0.83 $\pm$ 0.11 & 29.9 $\pm$ 30.9 & 0.86 $\pm$ 0.12 \\
		& (64x64x64) & & & & \\ \hline
	\end{tabular}
	\label{tab:3DUnet}
\end{subtable}
\label{tab:Unet}
\end{table} 

\begin{figure}[H]
	\centering
	\begin{minipage}{0.5\textwidth}
		\centering
		\includegraphics[width=1\linewidth, height=0.25\textheight]{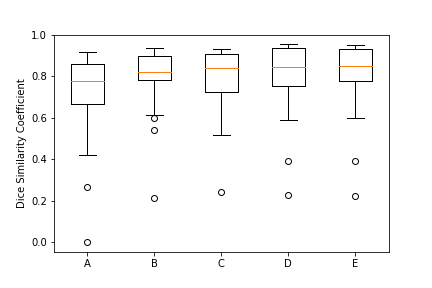}
		\label{fig:Boxplot_2D_DSC}
	\end{minipage}%
	\begin{minipage}{0.5\textwidth}
		\centering
		\includegraphics[width=1\linewidth, height=0.25\textheight]{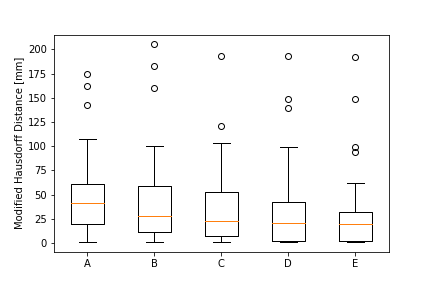}
		\label{fig:Boxplot_2D_MHD}
	\end{minipage}
	\begin{flushleft}
		\begin{minipage}{0.5\textwidth}
			\centering
			\includegraphics[width=1\linewidth, height=0.25\textheight]{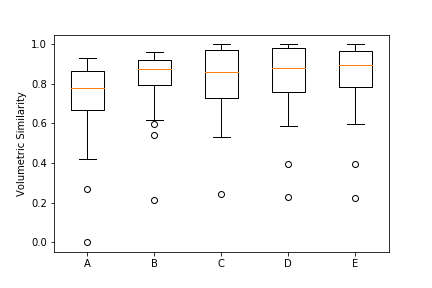}
			\label{fig:Boxplot_2D_VS}
		\end{minipage}%
	\end{flushleft}
	\caption{Boxplots of the vessel segmentation results obtained by training a 2D U-Net. The evaluation was performed by five experiments: A) without augmentation; B) augmented training data with Gaussian blurring; C) augmented training data with rotation and flipping; D) augmented training data with Gaussian blurring, rotation and flipping and E) trained on slices with augmented training data with Gaussian blurring, rotation and flipping.}
	\label{fig:boxplots_2D}
\end{figure}

In addition, Figure \ref{fig:boxplots_2D} shows the boxplots of the used evaluation metrics of the 2D U-Net segmentation results compared to the ground truth. According to this figure and a Wilcoxon signed-rank test, it is observed that the performance of the 2D U-Net improved by augmenting the training data (experiments (b)-(e)) compared to no augmentation (experiment (a)). This was observed from the DSC and VS of experiments (b)-(e), which were significantly higher compared to experiment (a) (p$<$0.05).

\begin{figure}[H]
	\centering
	\begin{minipage}{0.5\textwidth}
		\centering
		\includegraphics[width=1\linewidth, height=0.25\textheight]{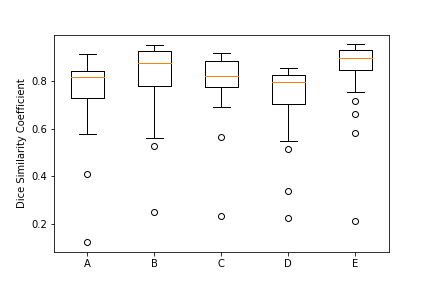}
		\label{fig:Boxplot_3D_DSC}
	\end{minipage}%
	\begin{minipage}{0.5\textwidth}
		\centering
		\includegraphics[width=1\linewidth, height=0.25\textheight]{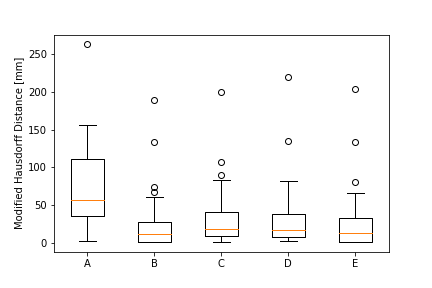}
		\label{fig:Boxplot_3D_MHD}
	\end{minipage}
	\begin{flushleft}
		\begin{minipage}{0.5\textwidth}
			\centering
			\includegraphics[width=1\linewidth, height=0.25\textheight]{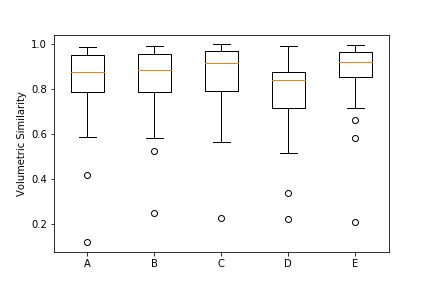}
			\label{fig:Boxplot_3D_VS}
		\end{minipage}%
	\end{flushleft}
	\caption{Boxplots of the vessel segmentation results obtained by training a 3D U-Net. The evaluation was performed by five experiments: A) without augmentation; B) augmented training data with Gaussian blurring; C) augmented training data with rotation and flipping; D) augmented training data with Gaussian blurring, rotation and flipping and E) trained on larger patches (64x64x64) with augmented training data with Gaussian blurring, rotation and flipping.}
	\label{fig:boxplots_3D}
\end{figure}

Figure \ref{fig:boxplots_3D} shows the boxplots of the DSC, MHD and VS computed from the 3D U-Net results. From this figure and a Wilcoxon signed-rank test it is also observed that the performance of the 3D U-Net was improved by augmenting the training data (experiments (b)-(e)) compared to no augmentation (experiment (a)). This can be observed from the DSC of experiments (b) (p=0.002) and (e) (p=$3.43*10^-8$), which were significantly higher compared to experiment (a). Additionally, the MHD of experiment (a) was significantly higher compared to all the other experiments (p$<$0.05). 

\begin{figure}[H]
	\begin{minipage}[b]{0.4\textwidth}
		\includegraphics[width=\textwidth]{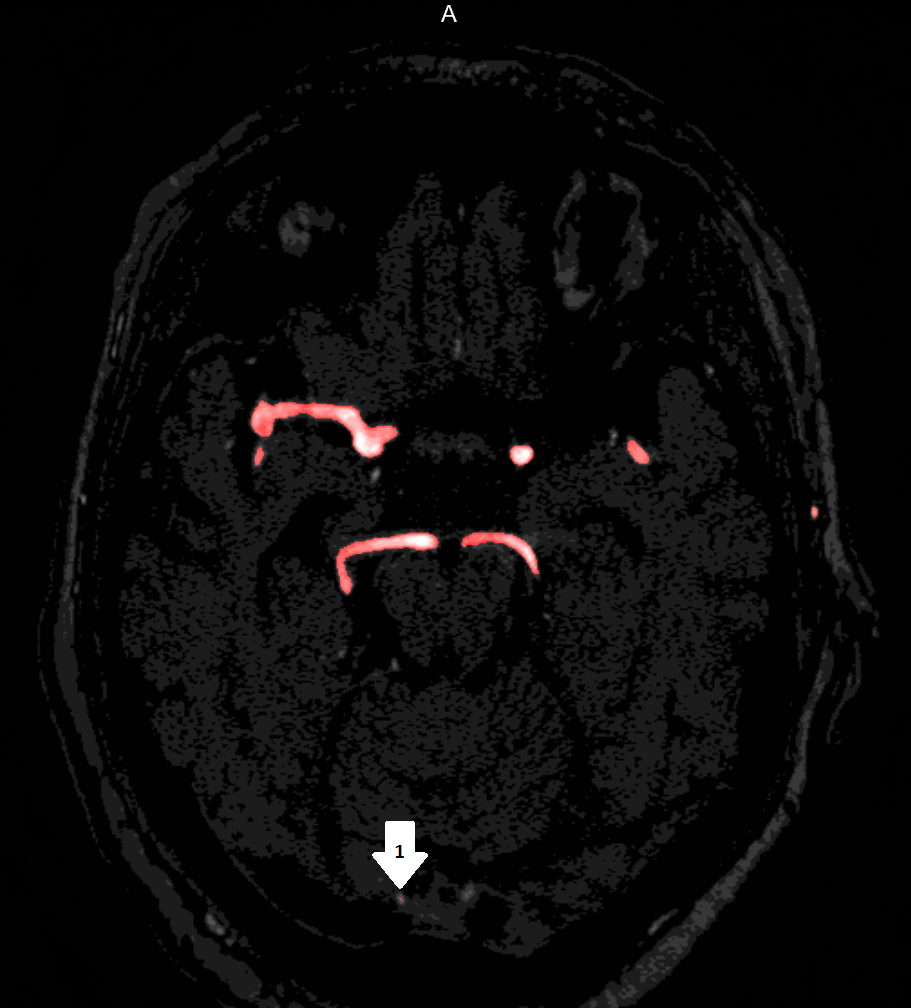}
		\subcaption{}
	\end{minipage}
	\hspace*{4cm}
	\begin{minipage}[b]{0.4\textwidth}
		\includegraphics[width=\textwidth]{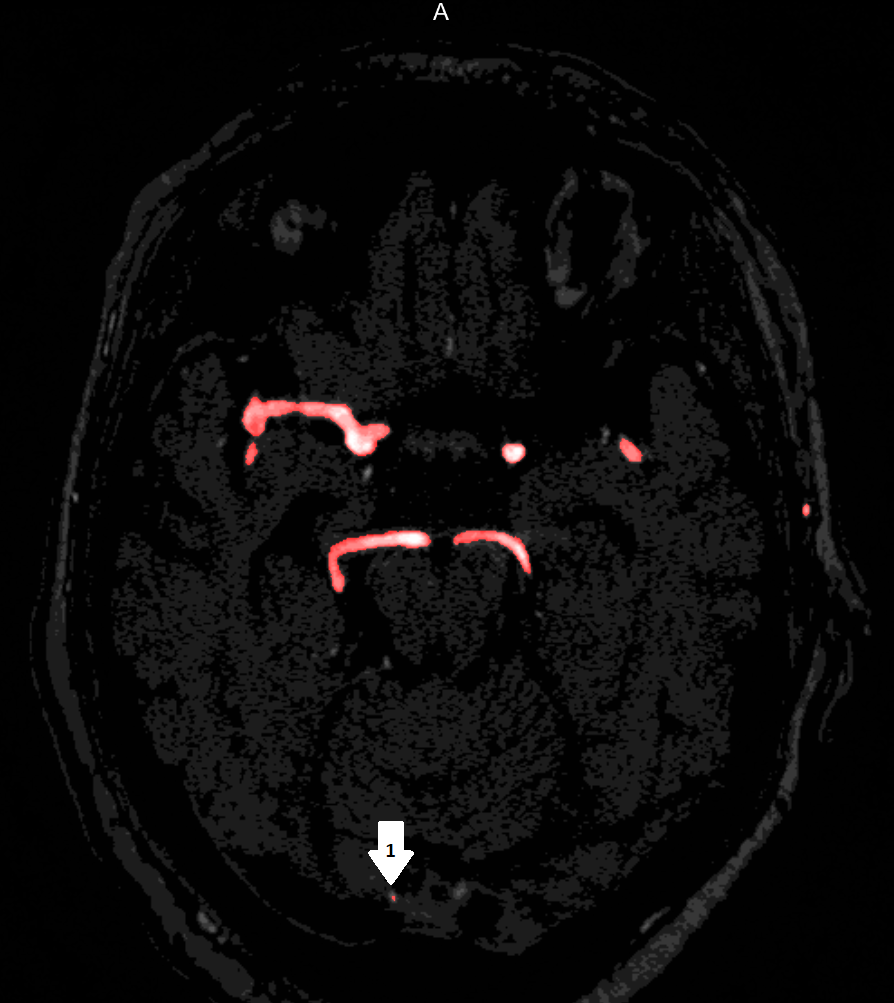}
		\subcaption{}
	\end{minipage}
	\vspace*{0.2cm}
	\caption{Example segmentation for one slice in the transverse plane of a TOF-MRA. (a) Ground truth segmentation. (b) Automatic segmentation resulted from the 3D U-Net trained on patches of size 64x64x64 voxels with Gaussian blur, rotation and flipping. Arrow 1 indicates a small oversegmentation in the automatic segmentation.}
	\label{fig:Example_1}
\end{figure}

\begin{figure}[H]
	\begin{minipage}[b]{0.4\textwidth}
		\includegraphics[width=\textwidth]{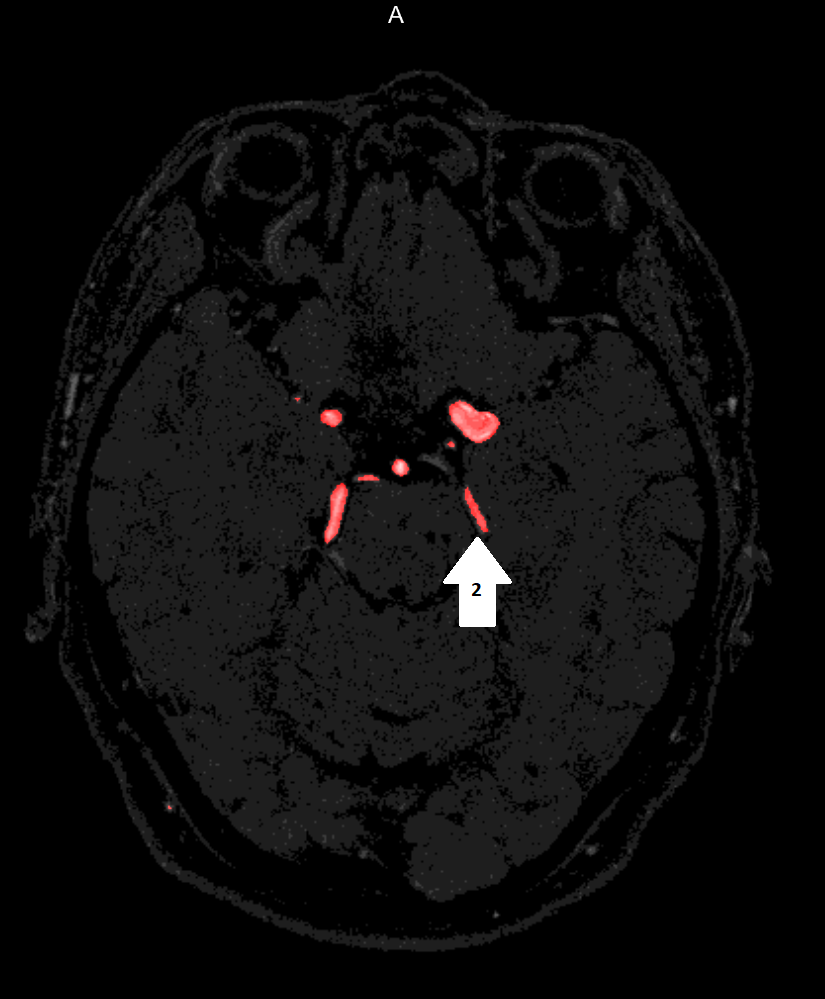}
		\subcaption{}
	\end{minipage}
	\hspace*{4cm}
	\begin{minipage}[b]{0.4\textwidth}
		\includegraphics[width=\textwidth]{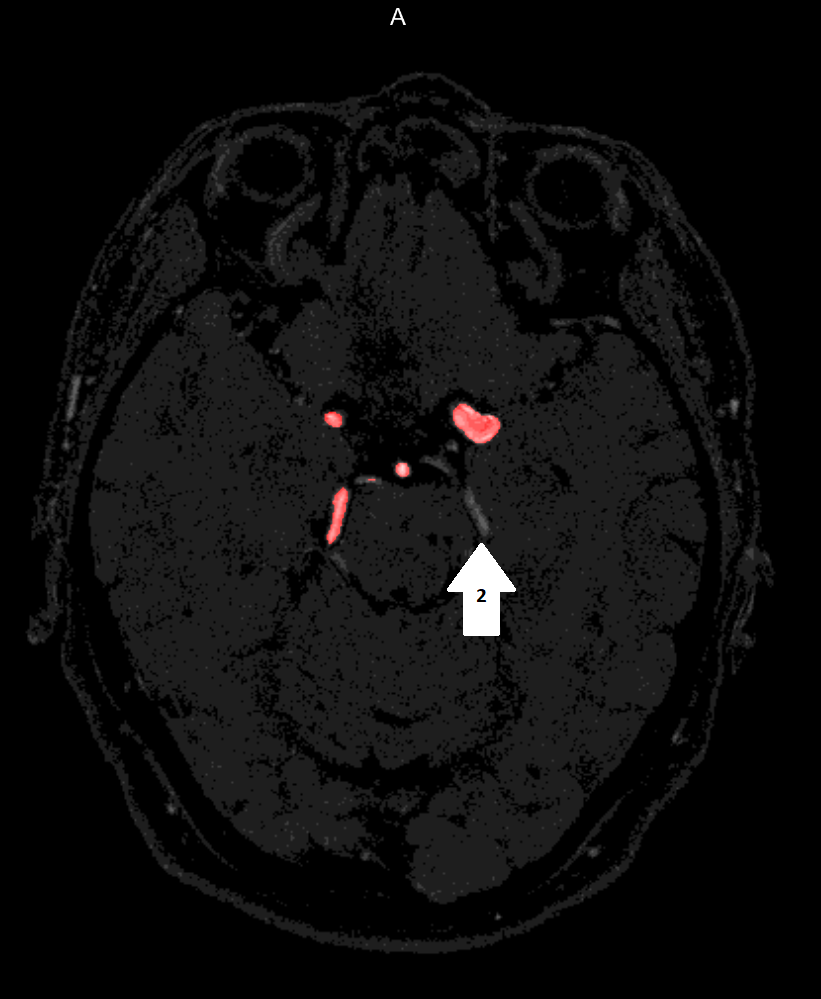}
		\subcaption{}
	\end{minipage}
	\vspace*{0.2cm}
	\caption{Example segmentation for one slice in the transverse plane of a TOF-MRA. (a) Ground truth segmentation. (b) Automatic segmentation resulted from the 3D U-Net trained on patches of size 64x64x64 voxels with Gaussian blur, rotation and flipping. Arrow 2 indicates an undersegmentation in the automatic segmentation.}
	\label{fig:Example_2}
\end{figure}

Figures \ref{fig:Example_1} and \ref{fig:Example_2} display the segmentation results of experiment (e), the optimally performing method. From Figure \ref{fig:Example_1}, no important differences were visually observed between the ground truth and automatically obtained segmentation result, as confirmed by the quantitative analysis. Only a small oversegmentation in the automatic segmentation was observed of a posterior cortical vein (arrow 1). In addition, Figure \ref{fig:Example_2} showed an undersegmentation in the automatic segmentation of the left posterior cerebral artery (arrow 2).

The optimum method for cerebrovascular segmentation was found to be the 3D U-Net trained on patches of size 64x64x64 voxels with all augmentation procedures, which resulted in a DSC of 0.83, MHD of 29.9 mm and VS of 0.86.

\section{Discussion} \label{Discussion}
Comparing the performance of the proposed deep learning experiments for vessel segmentation yielded some interesting results. This study showed that the automatic cerebrovascular segmentation can be accurately performed using a CNN with U-Net architecture. The performance of the U-Net can be improved with augmenting the training data. The optimum network for vessel segmentation was determined to be the 3D U-Net on patches of size 64x64x64 voxels and augmented by Gaussian blur, rotation and flipping.

As described in section \ref{Results}, all experiments performed with the proposed CNN with U-Net architecture resulted in good DSC scores ranging from 0.72 to 0.83. In general, this overlap measure was higher compared to the DSC of 0.74 reported in a study of Chen et al. (2017), which used a 3D convolutional autoencoder for vessel segmentation \cite{Chen2017}. Another CNN for vessel segmentation in TOF-MRA was proposed by a study of Phellan et al. (2017) and resulted in DSCs ranging from 0.764 to 0.786 depending on the number of images used for training \cite{Phellan2007}. On the contrary, the U-Net framework proposed by a study of Livne et al. (2019) showed higher overlap measure with a mean DSC of 0.88 \cite{Livne2019}. This could be caused by the larger patches this study used. The study of Livne et al. (2019) found an optimal patch size of 96x96 voxels \cite{Livne2019}. In our study, it was also found that the experiment with training on slices or larger patches gave the best results, as described in section \ref{Results}.

In general, the 2D U-Net performs better compared to the 3D U-Net for cerebrovascular segmentation except in the experiment with training on slices or larger patches (e). This may be caused by the complicated shape of the vessels in 3D, which makes it more difficult for the network to learn. 

As described in section \ref{Materials and Methods}, we performed experiments with Gaussian blur, rotation and flipping. The Gaussian blur could help the network to learn more robust features, by varying the contrast between the vessels and surrounding tissues. The rotation and flipping overcome positional biases. The combination of those augmentation techniques results in more diverse training data resulting in a better segmentation accuracy. This was also observed in section \ref{Results}, where it was described that experiments (d) and (e) gave the best results for both the 2D and 3D U-Net. This section also described that both the 2D and 3D U-Net performances were improved by augmenting the training data compared to no augmentation, which proves the importance of data augmentation to increase the diversity of the data without actually collecting new data.

Finally, for both the 2D and 3D U-Net, the best results were obtained by training on slices or larger patches (64x64x64 voxels) (experiment (e)). This was also reported in a study by Livne et al. (2019) \cite{Livne2019} and may be due to the larger patches providing a better representation of the small vessels in the full brain MRA and thereby improving the learning process of the vessel locations in the brain.

\subsection{Advantages and limitations} \label{Advantages and limitations}
The proposed vessel segmentation experiments have both advantages and limitations.

Firstly, the computation time of the algorithm is important in clinical use. As described in section \ref{Introduction}, this is one of the main reasons to provide an automatic vessel segmentation method. The trained U-Net can provide the vessel segmentations in the order of seconds per image.

Second, as described in section \ref{Materials and Methods}, the same MRAs were used for both the 2D and 3D experiments for vessel segmentation. In addition, the patches used for training the 2D network were of size 64x64 voxels and for the 3D network of size 16x16x16 voxels in order to train on the same number of voxels per patch in 2D and 3D. Those factors make it easier to compare the experiments performed in this study.

One main limitation of deep learning is the dependency on training data. This training data should be able to represent the unlabelled test data well enough to provide good results. In this study, the dataset consisted of patients with unruptured aneurysms. To obtain a more representative dataset for vessel segmentation, healthy patients and patients with other pathologies, such as vessels containing stenoses, occlusions or infarcts could be included. 

Another limitation of the vessel segmentation is the lack of a manually labeled vessel imaging dataset. One main advantage of the proposed vessel segmentation method was that an interactive vessel segmentation method (described in section \ref{Pre-processing}) was used for generating the ground truth labels. Manual annotations are labour and time intensive and this study showed that it is possible to produce a robust vessel segmentation without them. However, some small vessels were missed by the interactive ground truth generation technique. This was the main cause of the relatively high MHD results, described in section \ref{Results}. Further investigation into optimising the ground truth segmentation is warranted. 

Finally, as described in section \ref{Results}, the best vessel segmentation results were obtained by training a U-Net on slices in 2D or larger patches in 3D. However, the largest patches in 3D were of size 64x64x64 voxels as the patch size was limited due to memory constraints. This potentially reduced the performance of the 3D U-Net where more context might be needed. 

\subsection{Future work} \label{Future work}
As described in section \ref{Materials and Methods}, the data used for the vessel segmentation was randomly split into a training, validation and testing set. Cross-validation could be performed to ensure the robustness and generalization of the trained network. 

As this was a preliminary study, the test set used for the evaluation of the proposed vessel segmentation algorithm only contained 26 images, which is relatively small. Consequently, outlier images could have a large influence on the results. Future work could focus on using a larger dataset to evaluate the performance of the proposed segmentation method. For example, the full dataset provided by the Aneurysm Detection And segMentation (ADAM) challenge containing 113 sets of brain MR images for training and 142 sets for testing could be used \cite{Adam}.

Furthermore, future work could improve the ground truth used for the deep learning. With multiple medical experts, a systematic quantitative rating could be performed which includes the intra- and inter-rater variability and improves the ground truth segmentations.

In this study, only vessel segmentations generated by the U-Net architecture were evaluated. The U-Net architecture was chosen because of its prevalent and successful use in previous medical image segmentation problems. In future work, other network architectures for vessel segmentation could be investigated. However, due to the nature of this segmentation problem, no large improvements with respect to the U-Net performance are expected. In addition, a study of Livne et al. (2019) compared the performance of the U-Net to the performance of a U-Net with half of the convolutional layers. This resulted in comparable segmentation results and reduced the training time \cite{Livne2019}. Further research could focus on evaluating the half U-Net, or a U-Net with less parameters, for vessel segmentation and comparing the performance to the original U-Net performance as described in our study.

\section{Conclusion} \label{Conclusion}
In conclusion, our study found that a 3D U-Net trained on patches of size 64x64x64 voxels augmented using Gaussian blur, rotation and flipping performs optimally for vessel segmentation from TOF-MRAs.

\bibliography{Bibliography.tex}  
\bibliographystyle{spiebib} 

\end{document}